\documentclass[11pt,twoside]{article}

\usepackage{asp2006}
\usepackage{epsf}
\usepackage{psfig}
\usepackage{lscape}

\begin{document}
\title{Metallicity Evolution of Active Galactic Nuclei}
\author{Tohru Nagao}
\affil{National Astronomical Observatory of Japan,
       2-21-1 Osawa, Mitaka, Tokyo 181-8588, Japan;
       e-mail: tohru@optik.mtk.nao.ac.jp}
\author{Roberto Maiolino}
\affil{INAF, Osservatorio Astrofisico di Roma,
       Via di Frascati 33, 00040 Monte Porzio Catone, Italy}
\author{Alessandro Marconi}
\affil{Dipartimento di Astronomia e Scienza dello Spazio,
       Universit\`a di Firenze, Largo E. Fermi 2,
       50125 Firenze, Italy}

\begin{abstract} 
In this contribution we report our recent investigation of
the gas metallicity in active galactic nuclei and its
dependence on luminosity and redshift.
We compile large spectroscopic datasets of broad-line and narrow-line AGNs,
and compare them with the results of our photoionization models. 
Through the analysis of both the broad and the narrow emission-line 
regions, we find that: (1) for a given luminosity, there 
is no redshift dependence of the gas metallicity; (2)
for a given redshift, there is a significant correlation
between gas metallicity and luminosity;
(3) the luminosity-metallicity relation does no
show any evolution in the redshift range 
$2 \la z \la 4$.
\end{abstract}



\section{Introduction}

Understanding the galaxy evolution is one of the key
astrophysical topics of this decade. The gas metallicity
in galaxies provides important information because it
is a tracer of their star formation history.
However, the observation of 
faint emission lines (most of which are in the rest-frame optical range
and therefore shifted into the near infrared at high-z), associated with H~{\sc ii}
regions in star forming galaxies, is very difficult and time
consuming. Instead, AGNs exhibit bright
emission lines at rest-frame UV wavelengths, which can
be used to investigate the gas metallicity even in
high-$z$ objects .
In this contribution, we report our recent studies on
the gas metallicity of high-$z$ AGNs for both the broad-line 
region (BLR) and the narrow-line region (NLR).
Details are given in Nagao et al. (2006a), Nagao et al.
(2006b), and Maiolino et al. (2006).

\section{The BLR Metallicity}

\begin{figure}
  \plotone{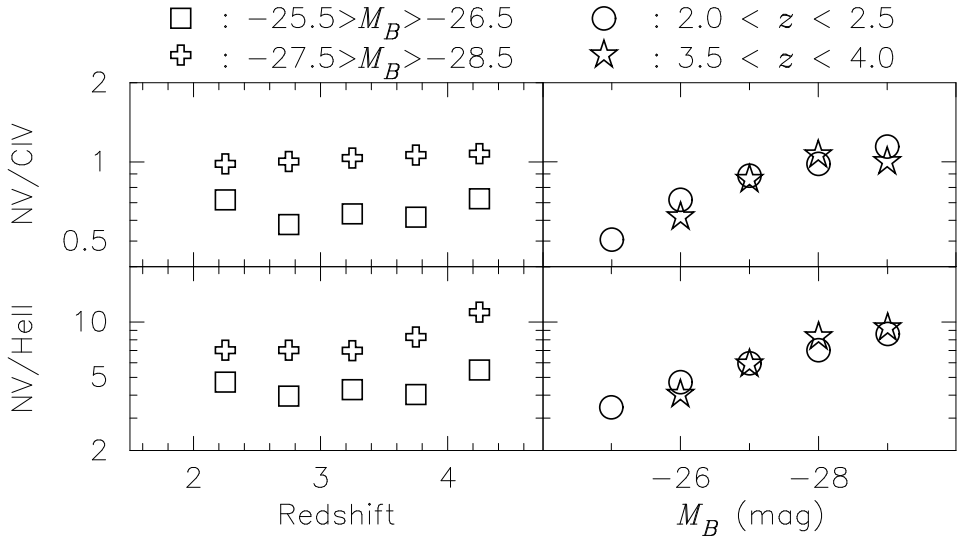}
  \caption{
     Metallicity-sensitive emission-line flux ratios,
     N{\sc v}/C{\sc iv} ($upper$) and N{\sc v}/He{\sc ii}
     ($lower$), measured in the composite spectra of the 
     SDSS quasars are shown as functions of redshift
     ($left$) and absolute $B$ magnitude ($right$).
  }
\end{figure}

The BLR metallicity in quasars is generally higher than
solar (e.g., Hamann \& Ferland 1992; Dietrich et al. 2003).
Some observations suggest that quasars at
higher-$z$ quasars show higher metallicity than lower-$z$
quasars (e.g., Hamann \& Ferland 1992). It is also 
recognized that the BLR metallicity tends to be higher in
more luminous quasars (e.g., Hamann \& Ferland 1993).
However, since higher luminosity quasars tend to be 
selectively observed at higher-$z$, the above relationes are degenerate.
More specifically, it was not clear 
whether the BLR metallicity depends primarily on the luminosity
or on the redshift.

To tackle this issue we focused on SDSS quasars.
We retrived 5344 
spectra of quasars at $2.0 \leq z \leq 4.5$ from the SDSS Data Release 2.
The quasars were divided into redshift and luminosity bins with
intervals $\Delta z = 0.5$ and $\Delta M_B = 1$ mag.
Then we made a composite spectrum for each ($z$, $M_B$)
bin, after correcting for the Galactic reddening, 
removing BAL quasars, and by applying an appropriate normalization.

The composite spectra show that there are significant 
correlations between the quasars luminosity and various
metallicity-sensitive emission-line flux ratios, such as
N{\sc v}/C{\sc iv} and N{\sc v}/He{\sc ii}, while there
are almost no correlations between the quasars redshift and
such emission-line flux ratios (Figure 1).
The correlation of line ratios with luminosity is
interpreted in terms of higher gas metallicity in more
luminous quasars. For a given quasar luminosity, there
is no significant metallicity evolution within the redshift
range $2.0 \leq z \leq 4.5$.
The absence of the metallicity evolution up to $z \sim 4.5$
may suggest that the active star-formation epoch of quasar
host galaxies occurred at $z \ga 7$, inferred by
timescale required for the enrichment of
some elements such as
C and Si (produced mainly by low- or intermediate-mass 
evolved stars).

\section{The NLR Metallicity}

\begin{figure}
  \plotone{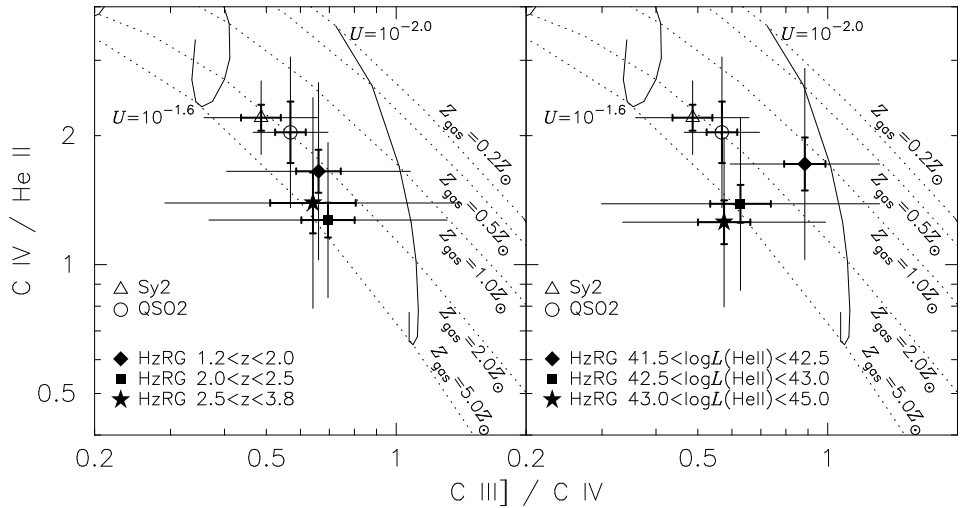}
  \caption{
    Averaged NLR flux ratios, compared with
    model predictions on the C{\sc iv}/He{\sc ii} versus
    C{\sc iii}]/C{\sc iv} diagram, for radio galaxies
    grouped in bins of redshift ($left$) and of He{\sc ii}
    luminosity ($right$). The dotted and solid lines denote
    the loci of predicted flux ratios at constant gas 
    density and at ionization parameter, respectively
    (assuming $n_{\rm H} = 10^5$ cm$^{-3}$).
    Thin error bars denote the RMS of the data distribution,
    and thick error bars denote the estimated errors on the
    averaged flux ratios.
  }
\end{figure}

The NLR metallicity offers information complementary to
the BLR metallicity since it traces spatial scales 
comparable with the host galaxies, although the measurements are
much more tough with respect to the BLR.
By using the N{\sc v}/C{\sc iv} flux ratio, De Breuck
(2000) claimed a NLR metallicity evolution of radio galaxies,
in their spectroscopic sample, from $z > 3$ to $z < 3$.
However, since the N{\sc v} emission becomes very
weak at low metallicities, only upper-limit fluxes
on N{\sc v} are available for the majority of high-$z$
radio galaxies, which makes the studies on the evolution of
the NLR metallicity difficult. Alternative diagnostics of
the NLR metallicity are required to make further progresses in the
uderstanding of the NLR metallicity at high-z.

We focused on C{\sc iv}, He{\sc ii} and C{\sc iii}], which
are the most strong NLR emission lines seen in the rest-frame
UV spectra. The C{\sc iv}/He{\sc ii} ratio is
expected to be sensitive to the NLR gas metallicity.
This is because the gas temperature decreases when the
metallicity increases and thus the collisional excitation of
C{\sc iv} is gradually suppressed, while the He{\sc ii} 
luminosity is basically proportional to the number of 
He$^{++}$ ionizing photons and thus rather insensitive to
the metallicity. The C{\sc iii}]/C{\sc iv} ratio is instead
sensitive to the ionization degree and therefore it is used to check
any dependence of C{\sc iv}/He{\sc ii} on the ionization
state of the gas. As a result, a diagnostic diagram that 
consists of these two flux ratios is quite useful
to estimate the NLR metallicity, without the need to rely
on weak emission lines
such as N{\sc v} (see also Groves et al. 2004).

We investigated the spectroscopic data of radio galaxies
at $1.2 \leq z \leq 3.8$ given by De Breuck et al. (2000),
and found that most of the flux ratios are inconsistent 
with shock models, suggesting that the NLRs of radio galaxies
are mainly photoionized. We then compared the
observed flux ratios with the results of our photoionization
model calculations obtained by using Cloudy (Ferland et al. 1998)
assuming one-zone, dust-free clouds. Note that NLR clouds 
emitting relatively high ionization lines are expected to be 
dust-free (see, e.g., Marconi et al. 1994; Nagao et al. 2003). 
The photoionization 
dust-free models provide two possible scenarios which are 
consistent with the observed data: low-density gas clouds 
($n_{\rm H} \la 10^3$ cm$^{-3}$) with a sub-solar metallicity 
($0.2 \la Z_{\rm gas}/Z_\odot \la 1.0$), or high-density gas 
clouds ($n_{\rm H} \sim 10^5$ cm$^{-3}$) with a wide range of 
gas metallicity ($0.2 \la Z_{\rm gas}/Z_\odot \la 5.0$).
Regardless of the specific interpretation, the observational
data do not show any evidence for a significant evolution 
of the gas metallicity in the NLRs within the redshift range 
$1.2 \leq z \leq 3.8$ (Figure 2). Instead, we found a trend 
for more luminous radio galaxies to have more metal-rich gas 
clouds, which is in agreement with the same finding in the
studies of the broad-line regions including ours (\S2).

%

\acknowledgements 
TN is a JSPS fellow. This work was partly supported by the 
Italian Space Agency (ASI) and the Italian Institute for 
Astrophysics (INAF).


\end{document}